\documentclass[aps,prc,twocolumn,showpacs,preprintnumbers,amsmath,amssymb,superscriptaddress]{revtex4-1}

\usepackage{graphicx}
\DeclareGraphicsExtensions{.pdf,.png,.jpg,.jpeg,.eps,.ps}
\usepackage{epsfig}
\usepackage{dcolumn}
\usepackage{bm}
\usepackage{epstopdf}
\usepackage{longtable}   
\usepackage{hyperref}
\hypersetup{
    citecolor=blue,
    colorlinks=true,
    linkcolor=blue,
    filecolor=blue,      
    urlcolor=blue,
    pdftitle={Overleaf Example},
    pdfpagemode=FullScreen,
    }
\usepackage{xr}
\usepackage[utf8]{inputenc}
\usepackage[T1]{fontenc}
\usepackage{lmodern}
\usepackage{csquotes}
\usepackage{tabularx}
\newcommand{\bn}{\begin{eqnarray}}
\newcommand{\en}{\end{eqnarray}}
\newcommand{\bc}{\begin{center}}
\newcommand{\ec}{\end{center}}
\newcommand{\bi}{\begin{itemize}}
\newcommand{\ei}{\end{itemize}}

\newcommand{\disregard}[1]{}

\newcommand{\bL}{\begin{Large}}
\newcommand{\eL}{\end{Large}}
\newcommand{\be}{\begin{equation}}  
\newcommand{\ee}{\end{equation}}
\newcommand{\ba}{\begin{eqnarray*}}
\newcommand{\ea}{\end{eqnarray*}}

\usepackage{tikz}

\newsavebox{\tmpstrikebox}
\newlength{\tmpstrikelen}

\begin{document}

\title{Rapid structural evolution of neutron-rich silicon isotopes toward \texorpdfstring{$N$}{N} = 28}
\author{G.~L.~Zimba}
\email{gzimba1@lsu.edu}
\affiliation{Facility for Rare Isotope Beams, Michigan State University, East Lansing, Michigan 48824, USA}
\affiliation{Department of Physics and Astronomy, Louisiana State University, Baton Rouge, Louisiana 70803, USA}

\author{H.~Iwasaki}
\affiliation{Facility for Rare Isotope Beams, Michigan State University, East Lansing, Michigan 48824, USA}
\affiliation{Department of Physics and Astronomy, Michigan State University, East Lansing, Michigan 48824, USA}

\author{B.~A.~Brown}
\affiliation{Facility for Rare Isotope Beams, Michigan State University, East Lansing, Michigan 48824, USA}
\affiliation{Department of Physics and Astronomy, Michigan State University, East Lansing, Michigan 48824, USA}

\author{Y.~Utsuno}
\affiliation{Advanced Science Research Center, Japan Atomic Energy Agency, Tokai, Ibaraki 319-1195, Japan}
\affiliation{Center for Nuclear Study, University of Tokyo, Hongo, Bunkyo-ku, Tokyo 113-0033, Japan}

\author{N. Shimizu}
\affiliation{Center for Computational Sciences, University of Tsukuba, Tsukuba 305-8577, Japan}

\author{N. Aoi}
\affiliation{Center for Nuclear Study, University of Tokyo, Hongo, Bunkyo-ku, Tokyo 113-0033, Japan}
\affiliation{Research Center for Nuclear Physics, Osaka University, Ibaraki, Osaka 567-0047, Japan}

\author{M.~Basson}
\affiliation{Facility for Rare Isotope Beams, Michigan State University, East Lansing, Michigan 48824, USA}
\affiliation{Department of Physics and Astronomy, Michigan State University, East Lansing, Michigan 48824, USA}

\author{T.~Beck}
\altaffiliation[Present address: ]{KU Leuven, Instituut voor Kern- en Stralingsfysica, 3001 Leuven, Belgium}
\affiliation{Facility for Rare Isotope Beams, Michigan State University, East Lansing, Michigan 48824, USA}

\author{J.~Chen}
\affiliation{Facility for Rare Isotope Beams, Michigan State University, East Lansing, Michigan 48824, USA}

\author{J.~Chung-Jung}
\affiliation{Facility for Rare Isotope Beams, Michigan State University, East Lansing, Michigan 48824, USA}
\affiliation{Department of Physics and Astronomy, Michigan State University, East Lansing, Michigan 48824, USA}

\author{A.~Douglas}
\affiliation{Facility for Rare Isotope Beams, Michigan State University, East Lansing, Michigan 48824, USA}
\affiliation{Department of Physics and Astronomy, Michigan State University, East Lansing, Michigan 48824, USA}

\author{A.~Ertoprak}
\affiliation{Physics Division, Argonne National Laboratory, Lemont, Illinois 60439, USA}

\author{P.~Farris}
\affiliation{Facility for Rare Isotope Beams, Michigan State University, East Lansing, Michigan 48824, USA}
\affiliation{Department of Physics and Astronomy, Michigan State University, East Lansing, Michigan 48824, USA}

\author{C.~Fransen}
\affiliation{Institut für Kernphysik der Universität zu Köln, D-50937 Köln, Germany}

\author{A.~Gade}
\affiliation{Facility for Rare Isotope Beams, Michigan State University, East Lansing, Michigan 48824, USA}
\affiliation{Department of Physics and Astronomy, Michigan State University, East Lansing, Michigan 48824, USA}

\author{S.~A. ~Gillespie}
\affiliation{Facility for Rare Isotope Beams, Michigan State University, East Lansing, Michigan 48824, USA}

\author{A.~Hill}
\affiliation{Facility for Rare Isotope Beams, Michigan State University, East Lansing, Michigan 48824, USA}
\affiliation{Department of Physics and Astronomy, Michigan State University, East Lansing, Michigan 48824, USA}

\author{K.~Kolos}
\affiliation{Lawrence Livermore National Laboratory, Livermore, California 94550, USA}

\author{D.~Lempke}
\affiliation{Facility for Rare Isotope Beams, Michigan State University, East Lansing, Michigan 48824, USA}
\affiliation{Department of Physics and Astronomy, Michigan State University, East Lansing, Michigan 48824, USA}

\author{I.~Lihtar}
\affiliation{Facility for Rare Isotope Beams, Michigan State University, East Lansing, Michigan 48824, USA}
\affiliation{Ru\dj{}er Bo\v{s}kovi\'{c}  Institute, HR-10002 Zagreb, Croatia}

\author{T.~Mijatovi\'c}
\affiliation{Ru\dj{}er Bo\v{s}kovi\'{c}  Institute, HR-10002 Zagreb, Croatia}

\author{S.~Neupane}
\affiliation{Lawrence Livermore National Laboratory, Livermore, California 94550, USA}

\author{S.~Noji}
\affiliation{Facility for Rare Isotope Beams, Michigan State University, East Lansing, Michigan 48824, USA}

\author{T.~Parry}
\affiliation{Facility for Rare Isotope Beams, Michigan State University, East Lansing, Michigan 48824, USA}

\author{A.~Revel}
\affiliation{Facility for Rare Isotope Beams, Michigan State University, East Lansing, Michigan 48824, USA}
\affiliation{Department of Physics and Astronomy, Michigan State University, East Lansing, Michigan 48824, USA}

\author{R.~Salinas}
\affiliation{Facility for Rare Isotope Beams, Michigan State University, East Lansing, Michigan 48824, USA}
\affiliation{Department of Physics and Astronomy, Michigan State University, East Lansing, Michigan 48824, USA}

\author{A.~Sanchez}
\affiliation{Facility for Rare Isotope Beams, Michigan State University, East Lansing, Michigan 48824, USA}
\affiliation{Department of Physics and Astronomy, Michigan State University, East Lansing, Michigan 48824, USA}

\author{E.~Schieb}
\affiliation{Facility for Rare Isotope Beams, Michigan State University, East Lansing, Michigan 48824, USA}
\affiliation{Department of Physics and Astronomy, Michigan State University, East Lansing, Michigan 48824, USA}

\author{D.~Weisshaar}
\affiliation{Facility for Rare Isotope Beams, Michigan State University, East Lansing, Michigan 48824, USA}

\author{Y.~Yamamoto}
\affiliation{Center for Nuclear Study, University of Tokyo, Hongo, Bunkyo-ku, Tokyo 113-0033, Japan}
\affiliation{Research Center for Nuclear Physics, Osaka University, Ibaraki, Osaka 567-0047, Japan}

\date{\today}

\begin{abstract}
Neutron-rich Si isotopes represent a unique case of shell evolution, exhibiting a robust shell closure at $N=20$
and pronounced quadrupole collectivity at $N = 28$. We report lifetime measurements of excited states in $^{40}$Si and the first simultaneous lifetime and heavy-ion inelastic-scattering measurements in $^{41}$Si. In $^{40}$Si, the extracted lifetimes for the $2_1^+$ and $(2_2^+)$ states indicate moderate quadrupole collectivity at $N=26$, together with signatures of triaxiality. In $^{41}$Si, two near-degenerate states at 570 and 658~keV exhibit comparable $B(E2)$ strengths as extracted from inelastic scattering, while the measured lifetimes indicate dominant $M1$ decays. The combined lifetime and inelastic-scattering results suggest an evolution toward oblate shape, consistent with large-scale shell-model predictions.

\end{abstract}


\maketitle
\setlength{\parskip}{0pt}

The evolution of shell structure and emergence of collectivity in nuclei far from stability are central themes in modern nuclear physics~\cite{RevModPhys.92.015002,SORLIN2008602,NOWACKI2021103866}. Experimental studies across the nuclear chart have shown that conventional magic numbers can weaken or disappear, while new magic numbers may emerge in exotic systems~\cite{PhysRevLett.97.112501,PhysRevLett.99.022503,PhysRevC.79.051303,HOFFMAN200917}. Shell evolution is now recognized as a widespread phenomenon at several traditional and subshell closures, including $N=8$, 20, 28, and 40. The proton--neutron tensor force and weak-binding effects have been proposed as mechanisms driving shell evolution and the resultant collectivity across the nuclear chart~\cite{PhysRevLett.104.012501,PhysRevC.86.051301,PhysRevC.76.054319}.

The silicon chain spans three neutron magic numbers: $N=8$ ($^{22}$Si), $N=20$ ($^{34}$Si), and $N=28$ ($^{42}$Si). The $N=8$ and $N=20$ closures correspond to $LS$-magic major-shell gaps~\cite{physics4020035}. The comparatively high $2_1^+$ excitation energies of $^{22}$Si and $^{34}$Si, relative to neighboring isotopes, indicate enhanced neutron shell stability~\cite{ffwt-n7yc,v57q-45qj,PhysRevLett.80.2081,BAUMANN1989458}. By contrast, the $N=28$ closure originates from a spin--orbit-driven shell gap associated with $jj$ coupling. The $Z=14$ proton subshell closure is also associated with a $jj$ magic number, potentially making $^{42}$Si doubly $jj$-closed with oblate-driving tendencies~\cite{physics4020035}. Surprisingly, $^{42}$Si has a low $2_1^+$ energy of 741~keV and a large $B(E2;0^+\rightarrow2_1^+)$ value of $500(90)~e^2\mathrm{fm}^4$, indicating pronounced quadrupole collectivity~\cite{b8xj-ycqk,PhysRevLett.122.222501,PhysRevLett.99.022503}. An open question is how quadrupole collectivity emerges toward $^{42}$Si, at the intersection of the weakened shell gap and oblate-driving effects associated with the doubly $jj$-closed configuration.

Shell-model calculations for silicon isotopes predict rapid structural evolution toward $N=28$, driven by proton $sd$-shell mixing due to excess $f_{7/2}$ neutrons, favoring oblate deformation~\cite{PhysRevLett.104.012501,PhysRevC.86.051301}. However, experimental information near $^{42}$Si remains scarce. In $^{40}$Si, spectroscopy is limited to the $2_1^+$ state at 986~keV and tentative $2_2^+$ and $4_1^+$ states at 1620 and 2524~keV, respectively~\cite{PhysRevLett.97.112501,PhysRevLett.109.182501}. The low-lying $2_2^+$ state is intriguing, since such states are often associated with triaxiality~\cite{PhysRevC.86.051301,osti_4091234}. Experimental information on $^{41}$Si is similarly sparse, with only two measurements reported to date~\cite{Gade2024,SOHLER2011417}. Systematics along the $N=27$ isotones and shell-model calculations suggest a $3/2^-$ ground state, contrary to the normal spherical-shell-model $7/2^-$ ordering~\cite{Gade2024,PhysRevLett.102.092501}. Several negative-parity states with tentative $1/2^-$, $3/2^-$, and $5/2^-$ assignments are also predicted to coexist at low excitation energies~\cite{Gade2024}.

Here, we report lifetime measurements of the $2_1^+$ and $(2_2^+)$ states in $^{40}$Si. We further report the first simultaneous determination of excited-state lifetimes and heavy-ion inelastic-scattering cross sections for low-lying states in $^{41}$Si.  The experimental results are compared with large-scale shell-model calculations. Our results provide new experimental evidence for rapid structural evolution in neutron-rich silicon isotopes, from triaxial features in $^{40}$Si at $N=26$ towards oblate deformation in $^{41}$Si as the doubly $jj$-closed $N=28$ system is approached.

The experiment was performed at the Facility for Rare Isotope Beams (FRIB) at Michigan State University. A secondary beam was produced by fragmentation of a 225~MeV/nucleon $^{48}$Ca primary beam on a $^{9}$Be target and separated with the Advanced Rare Isotope Separator (ARIS)~\cite{HAUSMANN2013349,PORTILLO2023151}. The cocktail beam consisted of $^{43}$P (78$\%$) and $^{41}$Si (18$\%$), with minor contributions from $^{42}$P, $^{43}$S, and $^{39}$Al. The $^{41}$Si beam energy at the S800 target position~\cite{BAZIN2003629,YURKON1999291} was 120~MeV/nucleon. The $^{41}$Si beam was used for inelastic-scattering and lifetime measurements of excited states in $^{41}$Si, whereas one-neutron removal from $^{41}$Si populated excited states in $^{40}$Si for lifetime measurements.

The TRIple PLunger for EXotic beams (TRIPLEX) device~\cite{IWASAKI2016123} held a 468-mg/cm$^{2}$ $^{9}$Be target upstream of a 2158-mg/cm$^{2}$ $^{181}$Ta degrader. Measurements were first carried out at a target--degrader separation of 20~mm and subsequently at 2~mm. The 20-mm foil separation allows inelastic-scattering events associated with the two foils to be identified separately, as short-lived states populated in the $^{9}$Be target are expected to decay before reaching the subsequent $^{181}$Ta foil. Large separation data were used to extract the absolute inelastic-scattering cross sections for $^{41}$Si and to provide the population ratios required for the lifetime analysis of $^{40,41}$Si~\cite{PhysRevC.94.024340}. The 2-mm setting was sensitive to excited-state lifetimes and was used for recoil-distance measurements of $^{40,41}$Si~\cite{DEWALD2012786,PhysRevLett.112.142502,PhysRevC.94.024340}.

The $\gamma$ rays emitted by the reaction residues were detected with the Gamma-Ray Energy Tracking In-beam Nuclear Array (GRETINA)~\cite{PASCHALIS201344,WEISSHAAR2017187} in coincidence with residues identified in the S800 spectrograph. GRETINA comprised 12 detector modules, with four detectors at $58^{\circ}$ and eight at $90^{\circ}$  with respect to the beam axis. The TRIPLEX plunger was positioned 20 cm upstream of the center of the GRETINA array. Reaction residues were identified event-by-event from time-of-flight and energy-loss information from the S800 spectrograph focal-plane detectors. The momentum vectors of the residues were extracted from the S800 ion-trajectory information and used for event-by-event Doppler correction.

\begin{figure}[h]
    \centering
   \includegraphics[width=0.48\textwidth,height=0.240\textheight]{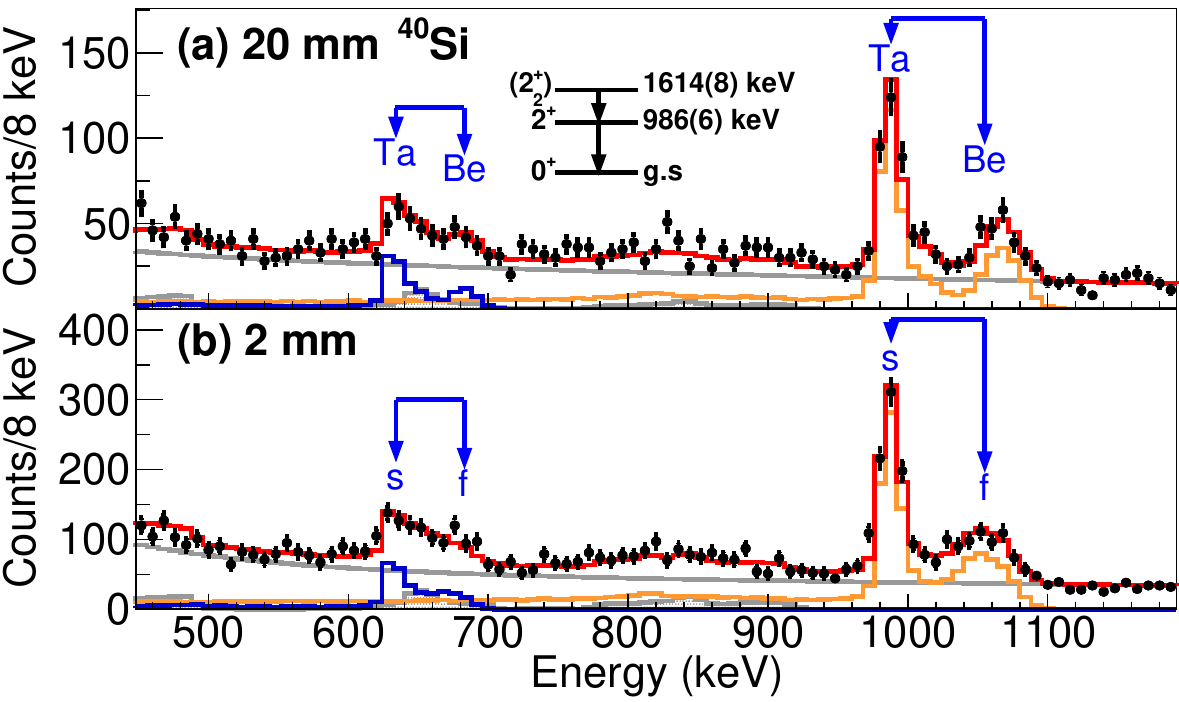}
\caption{Doppler-corrected $\gamma$-ray spectra of $^{40}$Si gated on laboratory angles below $45^{\circ}$ and optimized for decays following reactions in the Ta foil (slow component). Panels (a) and (b) show the 20- and 2-mm foil-separation data. Be-target and Ta-degrader contributions are indicated in (a); s and f denote the slow and fast components in panel (b). Data are shown as black points. Gray curves show the neutron-induced and exponential background components. Blue and orange curves show the simulated 628(4)- and 986(6)-keV transitions, respectively. The red curve the total fit.}
    \label{fig:Si40}
\end{figure}

We first present the results for $^{40}$Si, establishing the lifetimes of its excited states and providing a benchmark for the Doppler-correction and fitting procedure. Figure~\ref{fig:Si40} shows Doppler-corrected $\gamma$-ray spectra measured in the ($^{41}$Si,$^{40}$Si) channel. The prominent peaks at 628(4) and 986(6)~keV [Fig.~\ref{fig:Si40}(a,b)] correspond to transitions from the previously proposed $2_2^+$ and $2_1^+$ states of $^{40}$Si~\cite{PhysRevLett.109.182501,PhysRevLett.97.112501,CAMPBELL2007169}. From the 20-mm separation spectrum [Fig.~\ref{fig:Si40}(a)], the relative populations of the 986- and 1614-keV levels were determined to be 80(8)$\%$ and 20(8)$\%$, respectively.

Figure~\ref{fig:Si40}(b) shows the Doppler-corrected $\gamma$-ray spectrum of $^{40}$Si at a 2-mm foil separation. Lifetimes were extracted using the recoil-distance method~\cite{DEWALD2012786} by comparing the data with \textsc{Geant4}-based Monte Carlo simulations~\cite{AGOSTINELLI2003250}, following Refs.~\cite{PhysRevLett.112.142502,PhysRevC.94.024340}. Background contributions, including neutron-induced reactions, were constrained using laboratory-frame spectra and included in the Doppler-corrected fits.  The simulated spectra were compared with the data using $\chi^{2}$ minimization, with the lifetime parameter varied and the background and relative population yields constrained within their independently determined uncertainties.

The lifetime of the 1614-keV state was first obtained, and the lifetime of the 986-keV state was then extracted explicitly, accounting for feeding from the 1614-keV state. The best agreement yields $\tau = 13.2^{+1.5}_{-3.5}$ and $4.2^{+4.0}_{-2.8}$~ps for the 986- and 1614-keV states, respectively (see End Matter). Uncertainties are $1\sigma$ intervals including statistical and systematic contributions. Systematic uncertainties are dominated by feeding, through population intensities and lifetimes, and by background parameterization, through exponential amplitude and slope. The lifetime of the 1614-keV state is less well constrained owing to limited sensitivity and larger background uncertainties.

\begin{figure}[h]
   \centering
   \includegraphics[width=0.48\textwidth,height=0.240\textheight]{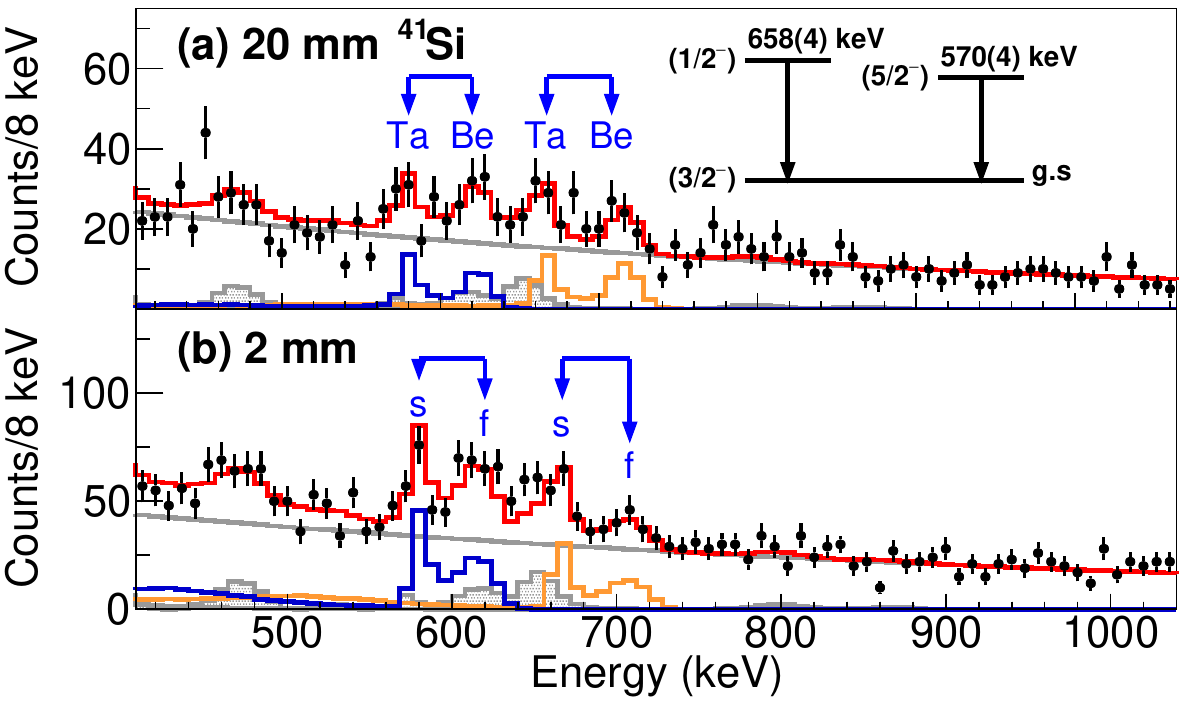}
\caption{Doppler-corrected $\gamma$-ray spectra of $^{41}$Si gated on laboratory angles below $45^{\circ}$ and optimized for decays following reactions in the Ta foil (slow component). Panels (a) and (b) show the 20- and 2-mm foil-separation data. Blue and orange curves show the simulated 570(4)- and 658(4)-keV transitions, respectively. The remaining symbols and curves are defined as in Fig.~\ref{fig:Si40}.}

  \label{fig:Si41}
\end{figure}

We now turn to $^{41}$Si, where inelastic-scattering cross sections are extracted from the 20-mm data and excited-state lifetimes from the 2-mm recoil-distance spectra. Figure~\ref{fig:Si41}(a) shows Doppler-corrected $\gamma$-ray spectra measured in coincidence with inelastically scattered $^{41}$Si residues at a 20-mm foil separation distance. Transitions at 570(4) and 658(4)~keV are observed, consistent with the 570(6)- and 649(6)-keV $\gamma$ rays reported in Ref.~\cite{Gade2024}. The extracted yields correspond to relative populations of 43(5)$\%$ and 57(5)$\%$ for the 570- and 658-keV states, respectively. The cross sections were integrated over laboratory scattering angles $\theta_{\mathrm{lab}}\leq 3^\circ$ and corrected for absolute GRETINA efficiency and S800 acceptance. For the 570-keV state, the cross sections are 1.4(3)~mb on the Be foil and 8.3(15)~mb on the Ta foil. For the 658-keV state, the corresponding values are 1.9(3)~mb and 10.9(18)~mb, respectively. Consistent with Ref.~\cite{Gade2024}, no statistically significant feeding transitions from higher-lying states are observed, therefore no feeding correction is applied. The quoted uncertainties include statistical uncertainties,
fit uncertainties, ambiguities in the determination of the S800 acceptance, and uncertainties in the absolute $\gamma$-ray efficiency, added in quadrature~\cite{REVEL2023137704,zp75-1ctq}.

The measured cross sections were compared with coupled-channel calculations performed using \textsc{Fresco}~\cite{THOMPSON1988167}, following the procedure established in Refs.~\cite{zp75-1ctq,REVEL2023137704,Wimmer2020}, to extract \textbf{$B(E2$$\uparrow;J_i\rightarrow J_f)$} values. The experimental angular resolution, angular straggling, and finite beam-energy spread were incorporated in the comparison with the measured cross sections. In the absence of elastic-scattering data for the present systems, a global optical potential derived from the complex $G$-matrix interaction CEG07~\cite{PhysRevC.85.044607,PhysRevC.80.044614} was employed. As this potential is available only for even-even nuclei, the $^{40}$Si+$^{181}$Ta and $^{40}$Si+$^{12}$C optical potentials were adopted to describe the $^{41}$Si+$^{181}$Ta and $^{41}$Si+$^{9}$Be systems, respectively.

The current \textsc{Fresco} calculations include two variables, the reduced transition matrix element $M(E2)$ which is directly related to the transition strength as $B(E\lambda;J_{i}\rightarrow J_{f})$ = $M(E\lambda)^{2}/(2J_{i} + 1)$, and the nuclear deformation length $\delta_N$, assuming dominant quadrupole $(l=2)$ excitation. The Be-foil cross sections were used to determine  $\delta_N$. Assuming excitation from the $3/2^-$ ground state to the $5/2^-$ and $1/2^-$ states corresponding to the 570- and 658-keV transitions, respectively, we obtained $\delta_N=0.45^{+0.05}_{-0.05}$~fm and $0.53^{+0.04}_{-0.05}$~fm. Using these $\delta_N$ values, the Ta-foil cross sections yield $B(E2$$\uparrow)$ values of $37^{+8}_{-9}$ and $48^{+10}_{-11}$~$e^{2}\mathrm{fm}^{4}$ for the 570- and 658-keV transitions, respectively. To investigate the sensitivity to the choice of optical potential, similar calculations were performed using the optical potential from $^{40}$Ar + $^{208}$Pb scattering at 41~MeV/nucleon~\cite{SUOMIJARVI1990369}, yielding $\delta_N=0.47^{+0.05}_{-0.05}$ and $0.56^{+0.04}_{-0.05}$~fm and corresponding $B(E2$$\uparrow)$ values of 41(10) and 53(11)~$e^{2}\mathrm{fm}^{4}$ for the 570- and 658-keV transitions. We therefore adopt averaged values of 39(10) and 50(15)~$e^{2}\mathrm{fm}^{4}$ for the 570- and 658-keV transitions, respectively, with uncertainties including both experimental contributions and optical-potential differences.

\begin{table}[t]
\centering
\caption{Experimental and shell-model results for states in $^{40}$Si and $^{41}$Si. MU and U-Si denote the SDPF-MU and SDPF-U-Si interactions. Excitation energies ($E_x$), lifetimes ($\tau$), $B(E2$$\uparrow;J_i\!\rightarrow\!J_f)$, and $B(M1$$\downarrow;J_i\!\rightarrow\!J_f)$ values are given in keV, ps, $e^2\mathrm{fm}^4$, and $\mu_N^2$, respectively. The $2_2^+$ assignment in $^{40}$Si and all $^{41}$Si spin-parity assignments are tentative. Calculations use effective charges $e_p=1.35e$ and $e_n=0.35e$.}
\label{tab:compare}
\setlength{\tabcolsep}{4.5pt}
\begin{tabular}{c c c c c}
\hline\hline
Nucl. & Obs. & Exp. & MU & U-Si \\
\hline \\[-10pt]

$^{40}$Si
& $E_x(2_1^+)$
& 986(6)
& 1121
& 803
\\[2pt]

& $E_x(2_2^+)$
& 1614(8)
& 1744
& 1251
\\[2pt]

& $\tau(2_1^+)$
& $13.2^{+1.5}_{-3.5}$
&
&
\\[2pt]

& $\tau(2_2^+)$
& $4.2^{+4.0}_{-2.8}$
&
&
\\[2pt]

& $B(E2;0^+\rightarrow 2_1^+)$
& $332^{+134}_{-43}$
& 323
& 271
\\[2pt]

& $B(E2;0^+\rightarrow2_2^+)$
&
& 104
& 70
\\[2pt]

& $B(M1;2^+_2\rightarrow2_1^+)$
& $<0.06^{+0.11}_{-0.03}$
& 0.03
& 0.02
\\[8pt]

$^{41}$Si
& $E_x(7/2^-_1)$
&
& 235
& 181
\\[2pt]

& $E_x(5/2^-_1)$
& 570(4)
& 517
& 785
\\[2pt]

& $E_x(1/2^-_1)$
& 658(4)
& 606
& 194
\\[2pt]

& $E_x(3/2^-_2)$
&
& 936
& 116
\\[2pt]

& $E_x(5/2^-_2)$
&
& 1143
& 1103
\\[2pt]

& $E_x(7/2^-_2)$
&
& 1194
& 846
\\[2pt]

& $\tau(5/2^-_1)$
& $8.1^{+2.7}_{-2.5}$
&
&
\\[2pt]

& $\tau(1/2^-_1)$
& $11.9^{+2.5}_{-2.3}$
&
&
\\[2pt]

& $B(E2;3/2^-_1\rightarrow7/2^-_1)$
&
& 142
& 12
\\[2pt]

& $B(E2;3/2^-_1\rightarrow5/2^-_1)$
& 39(10)
& 40
& 0.6
\\[2pt]

& $B(E2;3/2^-_1\rightarrow1/2^-_1)$
& 50(15)
& 67
& 59
\\[2pt]

& $B(E2;3/2^-_1\rightarrow3/2^-_2)$
&
&32
& 40
\\[2pt]

& $B(E2;3/2^-_1\rightarrow5/2^-_2)$
&
& 68
& 17
\\[2pt]

& $B(E2;3/2^-_1\rightarrow7/2^-_2)$
&
& 69
& 123
\\[2pt]

& $B(M1;7/2^-_1\rightarrow3/2^-_1)$
&
& 0.00
& 0.00
\\[2pt]

& $B(M1;5/2^-_1\rightarrow3/2^-_1)$
& $0.04^{+0.02}_{-0.01}$
& 0.11
& 0.05
\\[2pt]

& $B(M1;1/2^-_1\rightarrow3/2^-_1)$
& $0.014^{+0.004}_{-0.003}$
& 0.31
& 0.08
\\[2pt]

& $B(M1;3/2^-_2\rightarrow3/2^-_1)$
&
& 0.13
& 0.05
\\[2pt]

& $B(M1;5/2^-_2\rightarrow3/2^-_1)$
&
& 0.15
& 0.002
\\[2pt]

& $B(M1;7/2^-_2\rightarrow3/2^-_1)$
&
& 0.00
& 0.00
\\[2pt]
\hline\hline
\end{tabular}
\end{table}

The recoil-distance analysis of the 2-mm data [Fig.~\ref{fig:Si41}(b)] yields mean lifetimes of $\tau = 8.1^{+2.7}_{-2.5}$~ps for the 570-keV state and $\tau = 11.9^{+2.5}_{-2.3}$~ps for the 658-keV state. The presence of fast components in the Doppler-corrected spectra indicates the prompt nature of the 570- and 658-keV transitions. In the present experiment, the quantitative analysis is restricted to the two transitions discussed above, for which sufficient statistics are available (see End Matter). The present results are summarized in Table~\ref{tab:compare}.

We first discuss $^{40}$Si at $N=26$. The present experimental results are compared with large-scale shell-model calculations performed using the SDPF-MU~\cite{PhysRevC.86.051301} and SDPF-U-Si~\cite{PhysRevC.79.014310} interactions, both defined within the $sd$-$pf$ valence space. Following Refs.~\cite{PhysRevLett.97.112501,CAMPBELL2007169}, we interpret the 986-keV level as the $2_1^{+}$ state. The deduced $B(E2; 0^+\rightarrow 2_1^+)$ value for the $2_1^+$ state is $332^{+134}_{-43}~e^2\mathrm{fm}^4$, corresponding to $8.2^{+3.3}_{-1.1}$ W.u. The measured excitation energy and $B(E2)$ strength are well reproduced by the SDPF-MU interaction, while SDPF-U-Si predicts a comparable $B(E2$$\uparrow)$ value with a lower $2_1^+$ excitation energy. The extracted $B(E2)$ value indicates moderate quadrupole collectivity in $^{40}$Si at $N=26$, placing it on the evolutionary path toward the strong quadrupole collectivity observed in $^{42}$Si~\cite{b8xj-ycqk}.

For the state at 1614 keV, possible $0_2^+$, $2_2^+$, and $4_1^+$ assignments were considered. A $0_2^+$ or $4_1^+$ assignment is disfavored because the observed lifetime would correspond to $B(E2)\approx2\times10^3~e^2\mathrm{fm}^4$ ($\approx250$ W.u.) well above the recommended upper limit of 100 W.u. for $A\leq40$ nuclei~\cite{Endt1979}. Assuming pure $M1$ decay, the lifetime gives an upper limit of $B(M1)<0.06^{+0.11}_{-0.03}~\mu_N^2$. Adopting the recommended $E2$ upper limit as a partial contribution reduces the inferred $B(M1)$ strength to $0.03^{+0.07}_{-0.02}~\mu_N^2$, consistent with shell-model predictions in the range of $0.02$--$0.03~\mu_N^2$ from both SDPF-MU and SDPF-U-Si. These arguments favor a $2_2^+$ assignment of the state at 1614 keV.

The low excitation energy of the tentative $2_2^+$ state in $^{40}$Si is particularly striking, as it is amongst the lowest observed for nuclei with $A \leq 40$. Earlier SDPF-MU calculations~\cite{PhysRevC.86.051301} predict that the $2_2^+$ excitation energy along the silicon isotopic chain reaches a minimum at $N=26$, with the low energy interpreted as a signature of triaxiality. The present tentative $2_2^+$ assignment, together with the ratios $E(2_2^+)/E(2_1^+)=1.64$ and tentative $E(4_1^+)/E(2_1^+)=2.56$, and the measured $B(E2)$ strength, are fully consistent with the SDPF-MU calculations, supporting triaxial features for $^{40}$Si at $N=26$~\cite{PhysRevC.86.051301}.

We next discuss odd-mass $^{41}$Si at $N=27$. The excitation energies and $B(E2$$\uparrow)$ strengths are better reproduced by SDPF-MU than by SDPF-U-Si, favoring a low-lying structure in $^{41}$Si associated with the oblate band of $^{42}$Si rather than the prolate structure predicted by SDPF-U-Si~\cite{Gade2024}. Both interactions predict additional states above 1~MeV, but no statistically significant transitions are observed within the present sensitivity. The measured $B(E2)$ strengths imply pure-$E2$ partial lifetimes of 522 and 66~ps for the 570- and 658-keV decays, respectively; both exceed the measured lifetimes and indicate dominant $M1$ components (see End Matter). Assuming the level scheme of Ref.~\cite{Gade2024}, the 570-keV $(5/2^-)\rightarrow(3/2^-)$ and 658-keV $(1/2^-)\rightarrow(3/2^-)$ transitions yield $B(M1$$\downarrow)=0.04^{+0.02}_{-0.01}~\mu_N^2$ and $0.014^{+0.004}_{-0.003}~\mu_N^2$, respectively.  As observed for the low-lying $M1$ transitions in $^{37}$Si~\cite{PhysRevC.108.034304}, SDPF-MU overestimates the measured $B(M1)$ strengths, indicating that further refinement of the calculated wave functions may be required. Furthermore, neither the present data nor Ref.~\cite{Gade2024} provides evidence for the low-lying $7/2^-_1$ state predicted by SDPF-MU at 235~keV, with $B(E2;3/2^-\rightarrow7/2^-_1)=142~e^2\mathrm{fm}^4$. The non-observation of the predicted $7/2^{-}$ level is not conclusive, since its calculated mean lifetime of $16$~ns lies outside the region of sensitivity of the present setup.

\begin{figure}[h]
   \centering
   \includegraphics[width=0.48\textwidth,]{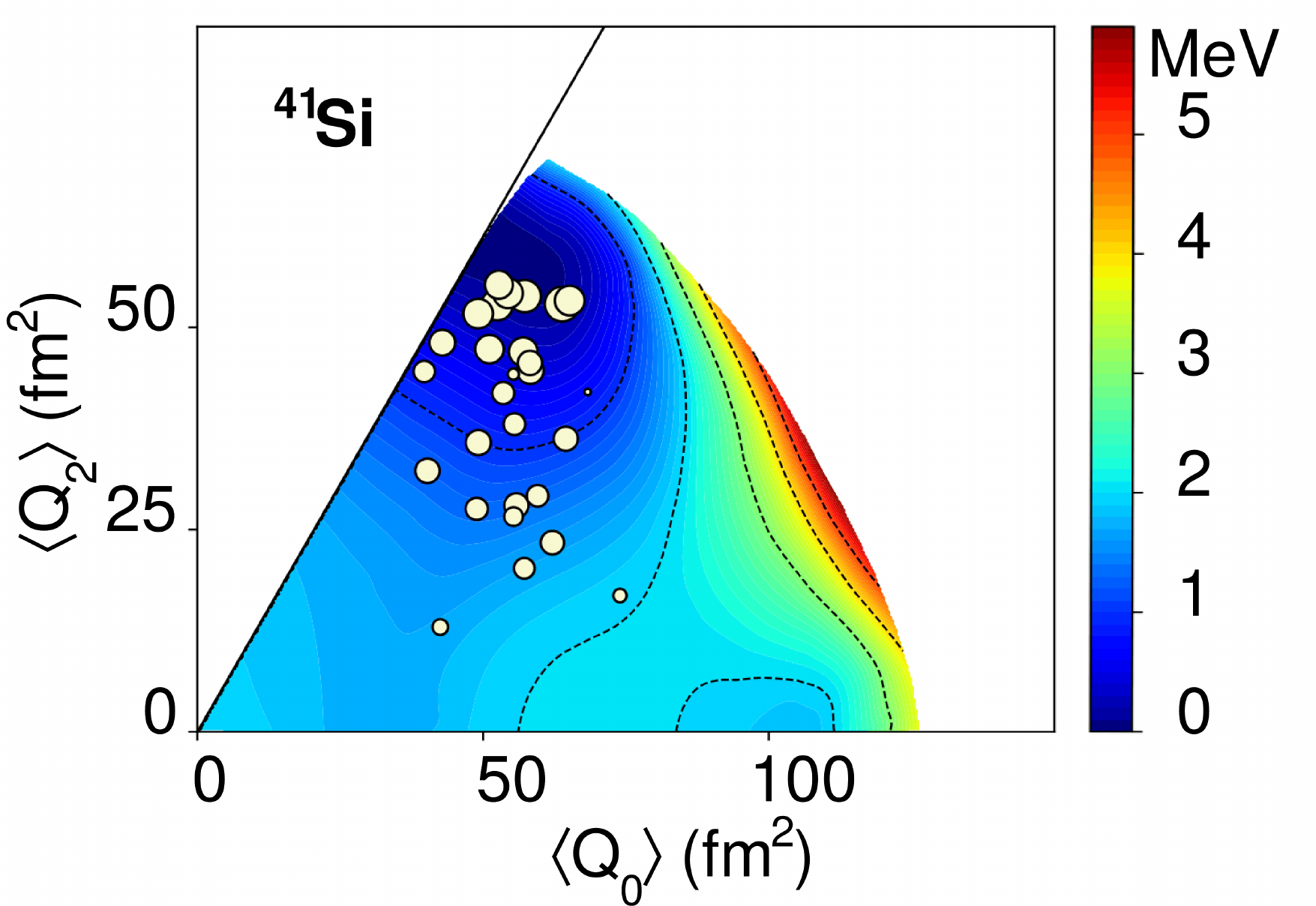}
 \caption{Potential-energy surface (PES) and T-plot for the $3/2^{-}$ ground state of $^{41}$Si as functions of the quadrupole moments. The energy relative to the minimum is shown by contour lines. Circles represent MCSM basis vectors, with sizes corresponding to overlap probabilities.}
  \label{fig:fig3}
\end{figure}

In the Nilsson model, the oblate deformation predicted for the doubly $jj$-closed nucleus $^{42}$Si is associated with the $1/2[330]$, $3/2[321]$, and $5/2[312]$ neutron orbitals, which are nearly degenerate at the Fermi surface~\cite{physics4020035,osti_4091234}. Importantly, at the doubly $jj$-closed $Z=14$ and $N=28$  configuration, the upsloping of the proton $5/2[202]$ and neutron $7/2[303]$ orbitals inhibit prolate deformation. The absence of this behavior in neutron-rich Mg and Ne is associated with the persistence of prolate deformation. For $^{41}$Si, a neutron-hole configuration involving the same oblate-driving negative-parity orbitals is expected to produce low-lying states with different $K$ values. As a reference for comparison with the present observations, the strong-coupling limit is considered for a $3/2^{-}$ ground state in $^{41}$Si. A $K=1/2$ sequence, for which the decoupling parameter of the $1/2[330]$ orbital ranges from $-4$ for the $f_{7/2}$ orbital to $-2$ for the $p_{3/2}$ orbital, places the $5/2^{-}$ state substantially above the $1/2^{-}$ state~\cite{BunkerReich1971,PhysRevC.100.014324}, whereas a $K=3/2$ band gives the regular $3/2^{-}$, $5/2^{-}$, $7/2^{-}$ ordering. In addition, a single $K=1/2$ ($K=3/2$) structure favors a dominant excitation to the $7/2^{-}$ ($5/2^{-}$) state, with other transitions carrying at most 40 - 50$\%$ of the leading $E2$ strength. The comparable excitation energies and $B(E2)$ strengths observed for the 570- and 658-keV states are not explained within a single $K$ quantum number, indicative of the presence or mixture of different configurations associated with the oblate deformation.

The potential-energy surface obtained with the SDPF-MU interaction provides
a consistent view, exhibiting a pronounced oblate minimum for
$^{41}$Si, as shown in Fig.~\ref{fig:fig3}. The circles indicate the intrinsic quadrupole moments of a given basis of the Monte Carlo Shell Model (MCSM) eigenstate for the $3/2^{-}$ ground state, created in a consistent way to the present theoretical calculations, and their size corresponds to the overlap probability \cite{PhysRevC.89.031301,Otsuka_2016}. These results suggest that excitations in $^{41}$Si arise from competing configurations, associated with the calculated minimum and the development of oblate deformation toward $N=28$.

In summary, we report lifetime measurements of excited states in $^{40}$Si and the first simultaneous lifetime and heavy-ion inelastic-scattering measurements in $^{41}$Si. In $^{40}$Si, the extracted $B(E2;0^+\rightarrow 2_1^+)$ value indicates moderate quadrupole collectivity at $N=26$, on the pathway to enhanced collectivity at $N=28$. The two near-degenerate states in $^{41}$Si at 570 and 658 keV exhibit comparable $B(E2$$\uparrow)$ strengths, while the measured lifetimes indicate dominant $M1$ decays. The present results provide new experimental evidence for rapid structural evolution from moderate quadrupole collectivity with triaxial features in $^{40}$Si toward oblate deformation in $^{41}$Si as the doubly $jj$-closed nucleus $^{42}$Si is approached, consistent with large-scale shell-model predictions.

\begin{acknowledgments}
This material is based upon work supported by the U.S. Department of Energy, Office of Science, Office of Nuclear Physics and used resources of the Facility for Rare Isotope Beams (FRIB) Operations, which is a DOE Office of Science User Facility under Award Number DE-SC0023633. The work is supported by the U.S. National Science Foundation (NSF) under Grant No. PHY-2110365. 
This work was performed under the auspices of the US Department of Energy (DOE) by the Lawrence Livermore National Laboratory (LLNL), USA under Contract No. DE-AC52-07NA27344. G.L.Z acknowledges support by the U.S. Department of Energy (DOE), Office of Science under Grant No.GR-00017863. Y.U. and N.S. acknowledge the support of the “Program for promoting research on the supercomputer Fugaku,” MEXT, Japan (Grant No. JPMXP1020230411), JST ERATO Grant No. JPMJER2304, Japan, and KAKENHI (Grants No. 25K00995 and No. 25K07330). C.F. acknowledges funding from the German Ministry of Education and Science, collaborative research project 05P2024 (ErUM-FSP T07), support code 05P24PK1. T.M. acknowledges the support of the Croatian Science Foundation under Project. No. UIP-2025-02-8168.

\end{acknowledgments}

\section*{End Matter} \label{End_Matter}

In the following, we provide additional data supporting the lifetime determination of the $(2_2^+)$ state in $^{40}$Si, as well as further details on the spectroscopy of $^{41}$Si carried out in the present work. We hope that the latter discussion, while less relevant to the main thrust of this Letter, will clarify how these supporting results connect to the main conclusions.

\section{A. Lifetime of the $2_{2}^{+}$ state in $^{40}$Si}
\begin{figure*}[]
  \centering
  \includegraphics[width=0.99\textwidth]{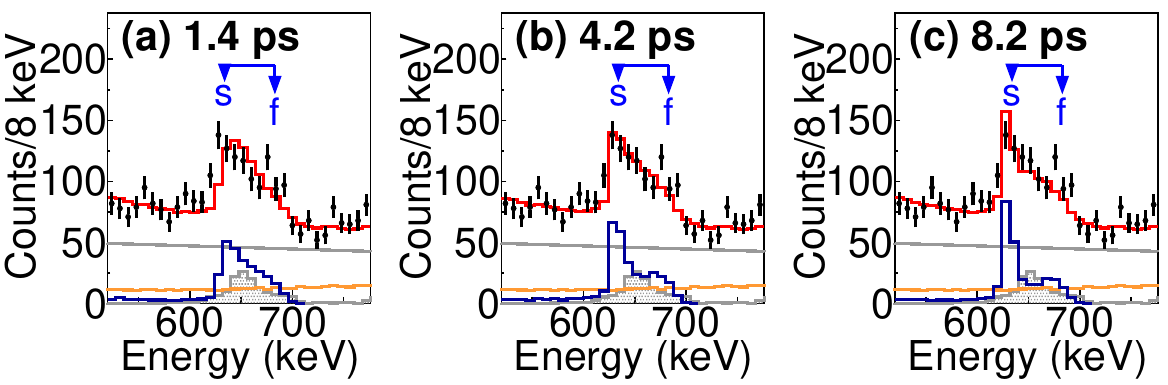}
\caption{Doppler-corrected $\gamma$-ray spectra of $^{40}$Si for the 628-keV transition at a target--degrader separation of 2~mm, gated on laboratory angles below $45^{\circ}$. Panels (a), (b), and (c) show comparisons with GEANT4 simulations for lifetimes of 1.4~ps, 4.2~ps, and 8.2~ps, respectively. Data are shown as black points, the neutron-induced and exponential background components by gray curves, the simulated 986(6)-keV transition by the orange curve, and the simulated 628(4)-keV transition by the blue curve. The fast (f) and slow (s) components are labeled accordingly. The red curve shows the total fit.}
 \label{fig:fig1_em}
\end{figure*}

The 1614-keV state in $^{40}$Si is associated with a lifetime of $\tau = 4.2^{+4.0}_{-2.8}$ ps, as determined from the spectral shapes of the fast and slow components of the 628-keV transition. To illustrate the sensitivity of the observed lineshape to the adopted lifetime, the 2-mm data are compared in Fig.~\ref{fig:fig1_em} with GEANT4 simulations for lifetimes of 1.4 ps [panel (a)], 4.2 ps [panel (b)], and 8.2 ps [panel (c)], corresponding to the lower $1\sigma$ bound, best-fit value, and upper $1\sigma$ bound, respectively. Background contributions from neutron-induced reactions and the 511-keV annihilation peak were evaluated in the laboratory frame and included in the Doppler-corrected fits through simulated response shapes, shown by the gray components in Fig.~\ref{fig:fig1_em}. The remaining smooth continuum was described by an exponential background, whose amplitude and slope were included as fit parameters and varied to estimate the uncertainty associated with the background treatment.

The $\tau = 1.4$~ps simulation fails to reproduce the observed lineshape, because a substantial fraction of decays occur within the Be target or the Ta degrader while the recoils are still traversing the foils. The 1.4-ps lifetime is substantially shorter than the estimated traversal time through the degrader foil, which is approximately 10~ps for the present beam energy. Consequently, $\gamma$ rays are emitted while the recoil ions still have a broad distribution of velocities, resulting in a Doppler-broadened lineshape rather than distinct fast and slow peaks [See Fig.~\ref{fig:fig1_em}(a)].

The $\tau = 4.2$~ps simulation provides the best overall description of the observed lineshape, reproducing the centroid, width, and relative contributions of the fast and slow components [See Fig.~\ref{fig:fig1_em}(b)]. In this regime, the lifetime is comparable to the characteristic traversal time through the degrader foil, such that decays occur before entering the degrader, while traversing the degrader, and after passing through the degrader, yielding the observed balance between the two components. The extracted value of $\tau = 4.2^{+4.0}_{-2.8}$~ps is therefore favored. The quoted uncertainties include statistical and systematic contributions added in quadrature. The systematic error arises primarily from the background parameterization, including the exponential amplitude and slope. No statistically significant evidence for additional feeding is observed within the present sensitivity.

The simulation with $\tau = 8.2$~ps presents more slow components corresponding to decays occurring after the degrader, resulting in a pronounced slow component with a strongly reduced fast component. The simulation clearly overestimates the slow component, as shown in Fig.~\ref{fig:fig1_em}(c). At this lifetime, corresponding to the upper $1\sigma$ bound of the 1614-keV state, the decay occurs predominantly after the recoil ions have traversed the degrader and reached their final velocity. Consequently, the Doppler correction becomes effective for a large fraction of events, leading to the formation of a distinct narrow peak associated with the slow component. Such a pronounced slow component is not supported by the experimental data.

\section{Spectroscopy of $^{41}$Si}

\begin{figure*}[]
  \centering
  \includegraphics[width=0.99\textwidth]{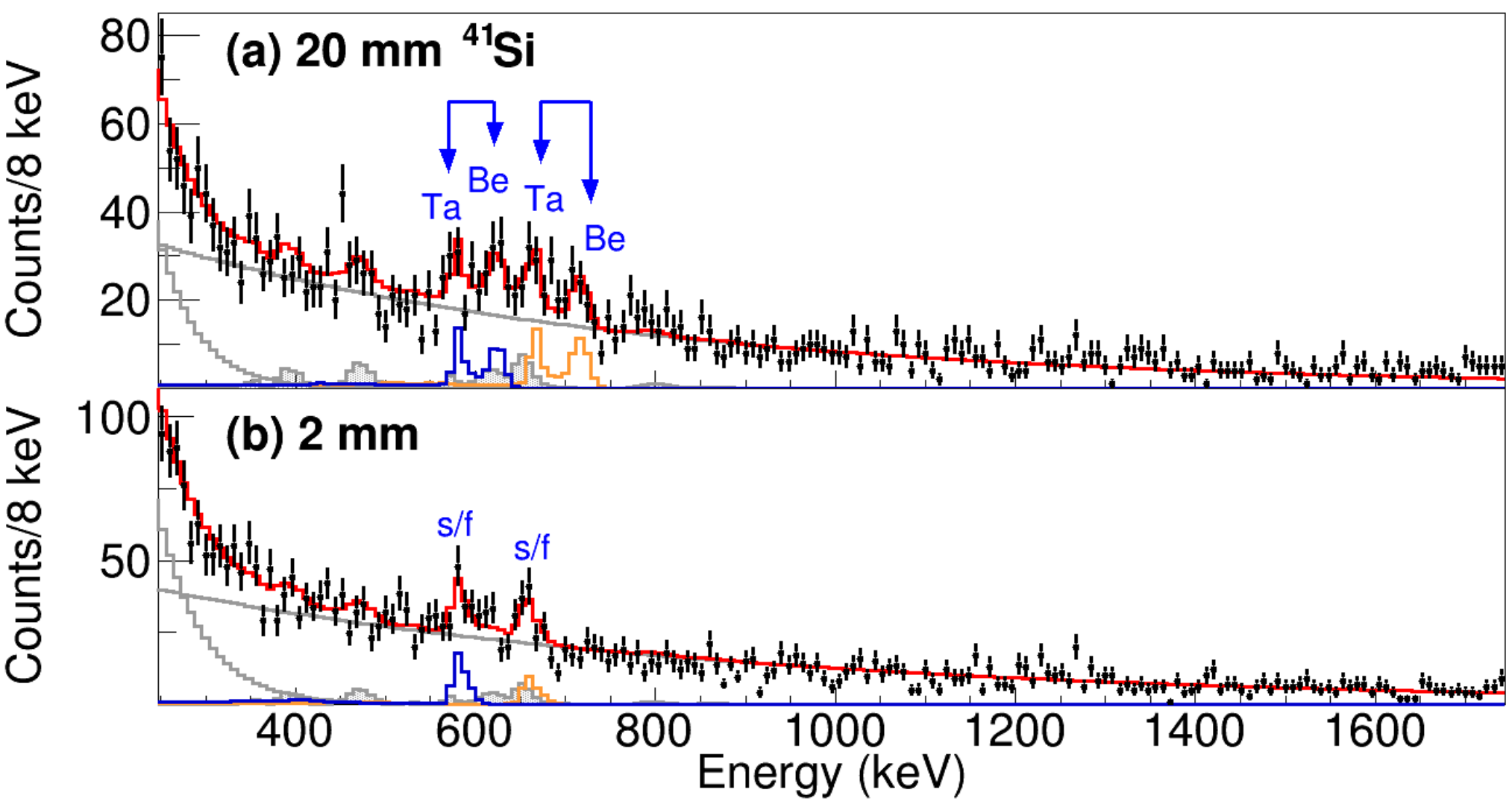}
  \caption{Doppler-corrected $\gamma$-ray spectra of $^{41}$Si. Panel (a) shows spectra obtained at a target--degrader separation of 20~mm, gated on laboratory angles below $45^{\circ}$ and optimized for the slow component. Contributions from reactions in the Be target and Ta degrader are indicated. Panel (b) shows spectra obtained at a separation of 2~mm, gated on laboratory angles above $55^{\circ}$, where the slow (s) and fast (f) components are not resolved and contribute to a single merged peak, indicated as s/f. Data are shown as black points. The neutron-induced background and the exponential background component are shown by the gray curves. Simulated 570(4)- and 658(4)-keV transitions are represented by the blue and orange curves, respectively. The red line shows the total fit.}

 \label{fig:fig2_em}
\end{figure*}

To verify that only two strong transitions contribute in the 600-keV region, we examine Doppler-corrected spectra obtained under conditions that maximize and minimize the separation of the components expected from the recoil-distance Doppler-shift method.

Firstly, for the 20-mm foil separation and laboratory angles below $45^\circ$ [Fig.~\ref{fig:fig2_em}(a)], distinct components associated with reactions in the Be target and Ta degrader remain evident. These arise from the different recoil velocities associated with reactions in the two foils, together with the long flight path between the target and degrader, which preserves the separation between the corresponding decay components.

On the other hand, for the data obtained with a foil separation of 2 mm and for laboratory angles above $55^\circ$ [Fig.~\ref{fig:fig2_em}(b)], the fast (f) and slow (s) components progressively overlap because of the reduced sensitivity to recoil-velocity differences at larger detector angles and the shorter flight path available for spatial separation of the decay locations. Under these conditions, the fast and slow components merge into a single peak for each transition.

The evolution from clearly separated components to merged single peaks follows the expected behavior of the recoil-distance Doppler-shift method and demonstrates that the peak splitting observed at forward angles arises from Doppler and lifetime effects rather than from additional unresolved transitions. The agreement of the peak centroids between the two conditions supports the interpretation that only two strong transitions, at 570 and 658 keV, contribute significantly within the present sensitivity.

While recent high-resolution measurements report a richer $\gamma$-ray structure in $^{41}$Si~\cite{Gade2024}, the present experiment is primarily sensitive to the two strongest transitions at 570 and 658~keV. Weak excess counts are visible at higher energies; however, their limited intensity and lack of resolved structure prevent a quantitative analysis. They are therefore not treated as identified feeding transitions and are not included explicitly in the lineshape fits or in the quoted uncertainties.

The measured lifetimes and extracted $B(E2$$\uparrow)$ values for the 570- and 658-keV transitions allow the total transition rate, $\lambda_{\mathrm{tot}}=\lambda_{M1}+\lambda_{E2}$, to be decomposed into its $M1$ and $E2$ components. Here, $\lambda_{\mathrm{tot}}$ is determined from the measured lifetime, while $\lambda_{E2}$ is deduced from the extracted $B(E2\uparrow)$ value. Assuming the proposed level scheme for $^{41}$Si [Fig.~4(a) of Ref.~\cite{Gade2024}], the 570-keV $(5/2^-)\rightarrow(3/2^-)$ transition yields $B(M1$$\downarrow)$ = $0.04^{+0.02}_{-0.01}\,\mu_N^2$, corresponding to $\lambda_{M1}/\lambda_{\mathrm{tot}}\approx98\%$, indicating an almost pure $M1$ decay. For the 658-keV $(1/2^-)\rightarrow(3/2^-)$ transition, $B(M1$$\downarrow)$ = $0.014^{+0.004}_{-0.003}~\mu_N^2$, with $\lambda_{M1}/\lambda_{\mathrm{tot}}\approx82\%$, again demonstrating clear $M1$ dominance.

Since the spin-parity assignments are not yet definitive, we also considered alternative spin assignments. In particular, the tentative $(5/2^-)$ and $(1/2^-)$ assignments of the 570- and 658-keV states were interchanged. The interchange modifies the $\lambda_{E2}$ value through the spin dependence of the inferred $B(E2$$\downarrow)$ values converted from the measured $B(E2$$\uparrow)$ strengths. In this scenario, the 570-keV transition corresponds to $\lambda_{M1}/\lambda_{\mathrm{tot}}\approx95\%$, while the 658-keV transition corresponds to $\lambda_{M1}/\lambda_{\mathrm{tot}}\approx94\%$. Therefore, the conclusion that both observed transitions are dominated by $M1$ decays remains unchanged.

\bibliography{Bibi}
\end{document}